# Information Fusion to Estimate Resilience of Dense Urban Neighborhoods

Anthony Palladino*[a], Elisa J. Bienenstock[b], Bradley M. West[a], Jake R. Nelson[b], Tony H. Grubesic[b]

[a] Boston Fusion Corp., 70 Westview Street, Suite 100, Lexington, MA, USA 02421;
[b] Arizona State University, 411 North Central Avenue, #750, Phoenix, AZ USA 85004

## ABSTRACT

Diverse sociocultural influences in rapidly growing dense urban areas may induce strain on civil services and reduce the resilience of those areas to exogenous and endogenous shocks. We present a novel approach with foundations in computer and social sciences, to estimate the resilience of dense urban areas at finer spatiotemporal scales compared to the state-of-the-art. We fuse multi-modal data sources to estimate resilience indicators from social science theory and leverage a structured ontology for factor combinations to enhance explainability. Estimates of destabilizing areas can improve the decision-making capabilities of civil governments by identifying critical areas needing increased social services.

## 1  INTRODUCTION

Two-thirds of the world's population is expected to live in urban areas by the year 2050. This rapid growth of dense neighborhoods with diverse sociocultural influences may strain civil services, e.g., law enforcement, health care, and waste management, reducing the resilience of these areas to exogenous and endogenous shocks. Aggregate statistics for large, sprawling, urban areas are likely to miss critical indicators.

To address these issues, we developed a novel approach that advances the state-of-the-art by (*i*) ascertaining the stability and resilience of densely populated areas at much finer geospatial and temporal scales, and (*ii*) basing that estimate on a solid social science foundation. We fuse multi-modal data sources to estimate the prevalence of indicators from social theories (e.g., social capital). Aggregate statistics from structured city-level data (censuses, surveys) provide a baseline for the entire city, while social media and news reports provide features at shorter time scales, and smaller geographic areas (i.e., neighborhoods, city blocks) introducing deviations from the baseline.

This social science foundation, combined with our structured ontology, leads to enhanced explainability of the stability assessment, increasing confidence in downstream decisions. Knowledge of destabilizing areas informs the decision-making process in civil governments by identifying neighborhoods in need of social services and civil control. Our computational framework estimates the capacity of a dense urban area to cope with stress and volatility, allowing governments, soldiers, humanitarian organizations, and private companies to better understand and manage local disturbances.

## 2  THEORETICAL FRAME: SOCIAL SCIENCE FOUNDATION

There is a need to bridge the gap between the computational and social sciences. Modern decision-makers require support from computational tools with near real-time feedback, however social science theories are still validated/quantified after lengthy surveys and structured interviews. We performed comprehensive research and built methods using open-source data and text analytics to allow data science to quantify social science theories at a rapid pace. As a first step, we considered the widely accepted concept of Social Capital as a measure of community health.

### 2.1  Social Capital: Definition in the Context of Resilience

The degree to which a community is vulnerable or resilient depends, in part, on how well community members can fend for themselves and take care of one another. Areas populated by "Hobbsian", atomistic, individualists are predicted to degrade quickly when government services cannot be provided. Communities composed of caring altruists, i.e., communities that exhibit a sense of "community" are predicted to be more resilient in the face of exogenous shocks. This is because, both from an individual and from an institutional perspective, communities comprised of people in the habit of working together and assisting one another under "normal" conditions are better able to coordinate during crises. From an individual vantage, people know where to seek and how to provide assistance and trust others enough to accept or give help. From an institutional perspective, people who need assistance know where to go to get support.

Social capital is a measure of social activities performed at one point in time, which, like economic capital investments, might payoff at some later point in time [1, 2, 3]. Individual investments include time spent together or favors done to assist friends and neighbors that ultimately garner networks of reciprocal support, which, in turn, build trust. Community level social capital emerges as people generalize trust beyond their direct contacts to care for public spaces, investing in public goods and "community" more generally.

As a social science concept, there is some controversy about how to properly measure social capital. Early studies of the phenomenon were not coordinated in the way they operationalized the concepts. For the purpose of this paper we are focused only on the literature related to community level social capital: evidence or the products of aggregated individual or institutional investment to improve the quality of community life. We ignore the academic controversy about whether social capital should be measured as activities that produce social capital, such as participation in voluntary organizations or reports of "helping" behavior, or the products of social capital, such as high levels of trust in people and government and well-kept neighborhoods, art, and locations for public gathering. Our objective is to remotely, without benefit of directly using questionnaires, measure social capital from available digital sources. In addition to measuring social capital generally, we also develop methods to identify whether indications of social capital observed reveal bonding or bridging social capital.

*Bonding social capital* refers to trust and reciprocity within homogeneous groups [2, 3, 4, 5, 6, 7, 8]. These dense social networks have strong ties with close connections. Bonding social capital benefits social support by creating strong in-group loyalty, trust, and social cohesion. This often comes at the cost of exaggerating sectarian divisions and perpetuating privilege. *Bridging social capital* refers to ties between members of different heterogeneous groups leading to linkages, and subsequent common interests and trust, among people from many social groups that comprise a community. These bridges create overlapping social circles that promote inclusion and provide more broad access to resources and opportunities. Bridging ties make sectarian fractures less likely, as more people become connected and have investments in relationships with members of the outgroup.

We also leverage *social anchor theory*, a specialty within the social capital literature that focuses specifically on how social institutions within a community serve as anchors which support the development of bridging social capital [9]. Schools, community centers, playgrounds, and sporting venues promote social capital insofar as they provide opportunities and locations (safe spaces) for people to gather. In communities where services are segregated there are fewer opportunities to generate positive intergroup bonds. In communities that promote intergroup activities bridging social capital is more likely. The prevalence of public art and physical structures such as schools, playgrounds, hospitals, community centers, churches, museums, and sporting arenas are measures of social capital. The degree to which these public goods are equally accessible to the many groups that occupy the community reveals whether the social capital is bridging or bonding.

## 3 MEASURING AND MODELING OBSERVABLES

We developed an *urban resilience ontology* that models the inter-relations between indicators and contains the prescription for combining them into a quantification of social theories as measures of resilience. In this paper, we describe our ontology using one such social theory: social capital. There are many indicators that we can use to model positive social capital: Trust, reciprocity, volunteerism, donations, midwifery, community groups, etc. Likewise, several indicators could be used to represent negative social capital: crime statistics by neighborhood, building quality, poverty, 311 calls, etc. Vacant space that is not owned or cared for and exhibits decay or disarray is another indicator of negative social capital.

### 3.1 Urban Resilience Ontology

Social capital cannot be directly measured. Instead, observable correlates and indicators must be defined to estimate social capital in a given neighborhood. We developed an urban resilience ontology, shown in Figure 1 (a), describing the relationship between observable indicators of social capital, like social structures and events, and factors affecting their contribution (either positive or negative) to social capital in a given area. Social structures include buildings and outdoor areas that provide value or services community like hospitals, places of worship, schools, parks, and fire stations. Social events bring members of the community together, like sporting events, fundraisers and farmers markets. The extent to which each structure or event contributes to increased or decreased social capital depends on several social factors shown in Figure 1 (b). Social factors are divided into two categories, access and capacity. Access addresses which groups within the community can attend events, or use services provided in social buildings. It also addresses distance, community members living close to a social event or social structure are more likely to benefit from it than members living farther away. Capacity refers to the number of people that can attend an event, or occupy a building (e.g., number of seats in a place of worship, or number of beds in hospital).

The Urban Resilience Ontology defines how to combine indicators of social capital and factors affecting them. For example, many equally distributed mosques within a predominately Muslim neighborhood would contribute to greater social capital than the same number and distribution of churches in this neighborhood, which would either go unused or serve a small number of people, leaving some unserved. This is especially true if there were also a paucity of Mosques in this area. Likewise, one mosque equal to the capacity of several smaller, uniformly distributed mosques, would provide lower social capital because neighborhood wide access would be lower, people closer to the Mosque would have privileged access. Similarly, hospitals with a religious affiliation that restrict access to their members provide greater social capital to members of the same religion than members of other religions, in a manner that can negatively impact community level social cohesion.

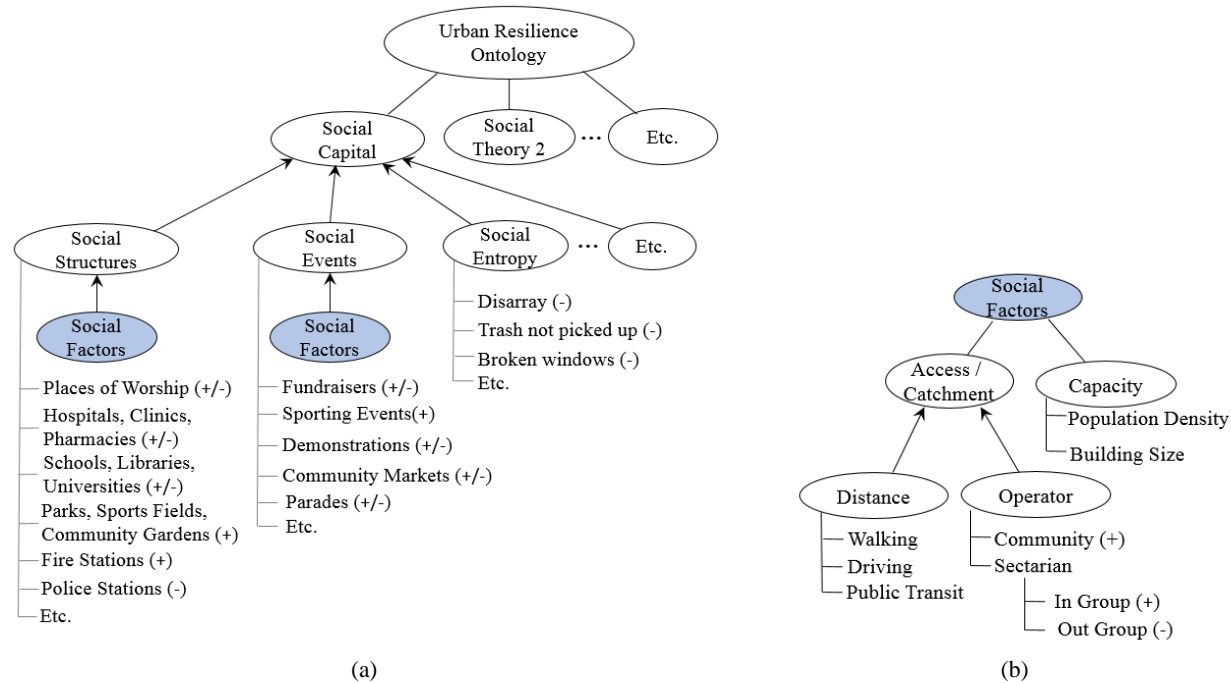

Figure 1: (a) Schematic overview of the Urban Resilience Ontology for the first social theory (*social capital*), and (b) social factors included in the quantification of social capital from social structures and events.

### 3.2  Geospatial Data Sources

To identify neighborhood boundaries and the geographic distribution of social structures defined in the Urban Resilience Ontology, we obtained data for Jakarta, Indonesia from OpenStreetMap [10]. OpenStreetMap is a community driven spatial database that contains world-wide information about locations of roads, restaurants, subway stations, hospitals, movie theaters, etc.

The spatial clustering or density of social structures on its own lacks sufficient meaning as an indicator of social capital. We must measure the density of social structures relative to the local population. Most population data available from censuses are only available at the city level. Worldpop.org, however, uses aerial and satellite imagery and machine learning, to estimate urban populations with 100m × 100m resolution, as shown in Figure 2 [11].

### 3.3 Kernel Density Layers

Kernel density estimation (KDE) is a set of techniques applied to point or line data sets to show a density value weighted by the distance from some starting feature—the output being a raster surface [12]. KDE is based on extensive spatial data, in other words spatial data that is dependent on the spatial unit (e.g., population totals, socio-economic status, or health metrics). By using variables within kernel density estimators, it is possible to obtain measures of spatial accessibility. For example, density surfaces showing the supply and demand for a particular service can be calculated and compared across a landscape to determine who needs access and whether there is adequate access to those locations [13].

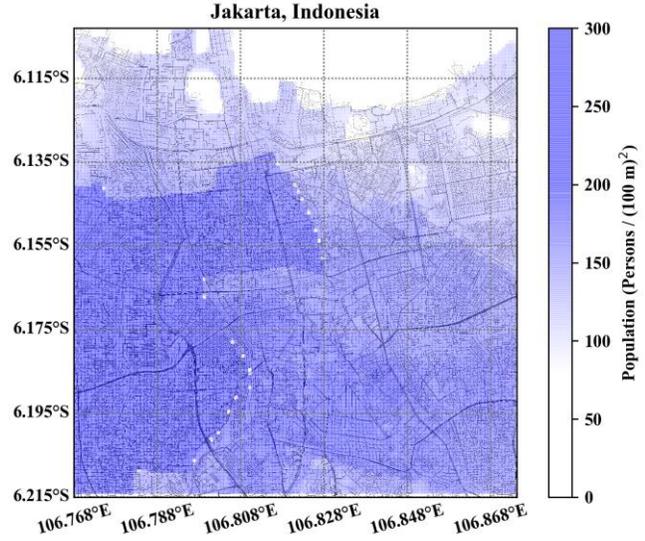

Figure 2: Population of Jakarta, Indonesia with 100m × 100m resolution.

Kernel density estimation is advantageous for analyzing the potential area of influence that a given location or geographic feature has on the surrounding landscape [14]. That is, kernel density can be used to assess a sphere of influence based on geographic proximity. For example, a kernel density surfaces have been used to determine the school that students are likely to attend based on the known locations of individuals who attend the school and the distance that each of those student locations are to the school [15]. The location and distances are used together in a weighting procedure to derive and associated "intensity" surface of geographic influence and feature membership.

The kernel density surface can also be advantageous in a number of other research contexts including the identification of crime [16] and traffic accident hotspots [17]. KDE has been used extensively for modeling the location of medical and healthcare facilities to determine how to approximate ambulance service areas for different medical facilities or, similar to other analyses, determine accessibility [13, 18, 19]. The power of the kernel density method is the ability to take into account both the density of the population and the distance to the facility in question. This also has the potential to provide an indication of resource availability in the context of resilience. For example, areas with high population but low access might be more vulnerable than those areas with higher levels of accessibility. However, there should still be some consideration for the demand and supply of the area. One example by Guagliardo [13] showed that a high density of hospitals does not necessarily translate to accessibility when combined with at-risk population locations. Taking into account two kernel density surfaces, one for the supply of physicians and one for the population of patients, [13] showed a heterogeneous landscape of high and low accessibility. When used as a method for defining the geographic influence of a feature on the surrounding landscape, kernel density surfaces can, in many ways, be considered a catchment area analysis (see below).

We defined a kernel density map layer for each category of social structure defined within the Urban Resilience Ontology. For example, the *places of worship* layer (Figure 3 (b)) includes the geographical distribution of mosques, churches, and temples in Jakarta. We developed our test prototype using a simple kernel: a 2-dimensional Gaussian distribution centered on each social structure in the layer. The amplitude, $A$, of the distribution is defined by

$$A = w * \frac{C}{P(A_c)}$$

where $C$ is the capacity of the building (e.g., estimated by square footage, the number of hospital beds, etc.), $w$ is a weight defined by our ontology ranging from -1 to 1 that describes whether the structure contributes positively or negatively to the local social capital, $P(A_c)$ is the integrated population within the building's catchment area, $A_c$ (described below).

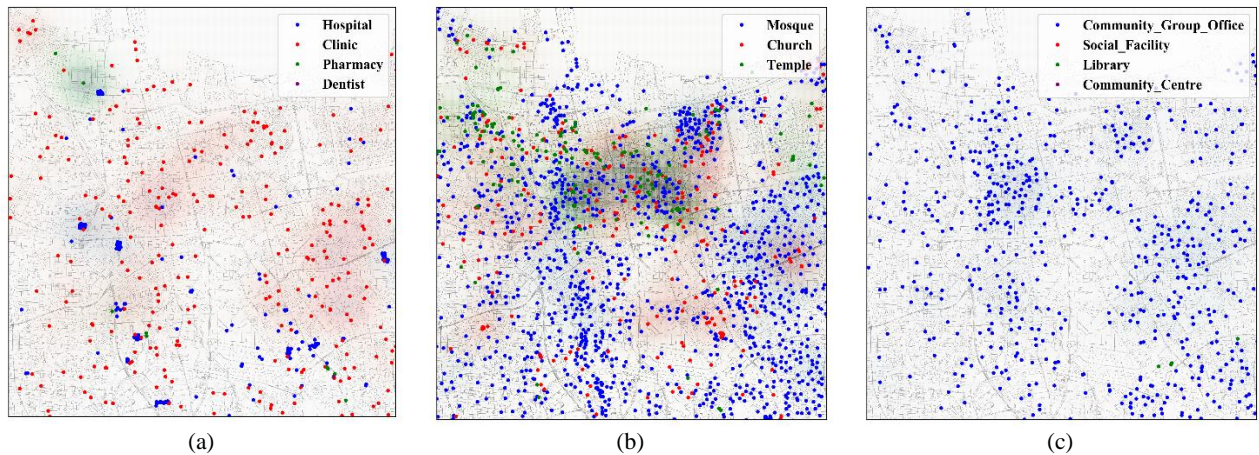

Figure 3: Kernel density maps for medical buildings (a), places of worship (b), and community buildings (c).

**Catchment Areas**

A catchment area, in the traditional sense, is the location where water sources eventually end up. This is usually a function of hydrography, geography and topology. Catchment areas can also refer to a geographic area that bounds a population that receives some sort of local service. In the latter case, examples include schools and the delineation of where students live, or the area that hospitals receive most of their patients from. Catchment area analysis has been extended to include the delineation of boundaries for more than just schools and hospitals. Recent studies have extended problems to include retail store locations [20], bike share access [21], and food store accessibility [22]. Researchers investigating catchment areas are often interested in determining who is served (in terms of socio-economic or demographic characteristics), the size of the catchment area, and relatedly, the ability of the service to meet the demand of the population in the catchment.

Access, broadly defined, is the ability of an individual to reach a particular location. It is often conceptualized as a measure of distance, where researchers place and upper bound on how far away a particular destination is from a particular population. In some work this distance is modeled using a simple radius of influence around each unit of analysis [23]. Other studies have used travel-time based on road networks combined with information on traffic flows [24, 25]. When calculating catchment based on access to hospitals, previous work has used travel time to hospitals based on a traffic Application Programming Interface (API), population data for 58 Chinese sub-districts, and evaluated their model for different travel thresholds [24].

One of the first questions to answer when defining a catchment area is determining which groups of people and/or locations have access. This necessarily involves the evaluation of who can realistically get to the location in question. Often, these studies are on underserved communities; hence the abundance of extant research on public transit access. To a certain degree these analyses can be thought of as a mini-catchment area for the transit stops [26]. When evaluating transit stop catchment, researchers determine access in terms of walkability. For example, Biba et al. [27] evaluate access to public transit stops using the size of the population in an analysis zone with access to transit, the total length of the street network, the length of the street network within walking distance to the transit zone, and the total population of the entire transit zone. Findings from other research suggests that at a distance of about 300 feet from the transit station, usage by those who walk will begin to decrease and at a half-mile the usage disappears altogether. With this in mind, researchers calculate the distance from each land-parcel centroid to the transit station to create buffers that encompass the likely access/catchment for each transit station. After identifying these buffer zones, evaluating associated statistics on who (and how many) is served is possible with census data. A similar approach is taken by [28] and [18] where the former develops the area public transit access defined by degree of convenience in getting from one traffic zone to another and the latter evaluates diurnal transit access patterns using differences in ridership for different times of the day.

A next step in determining catchment includes the evaluation of accessibility to destinations based on public transportation network destinations. One of the major goals of any transportation network is the ability to get people to where they need to go [29]. This has been evaluated with retail store locations [20], libraries, and supermarkets [30]. The determination of who has accessibility to certain amenities by way of public transportation (or other transportations methods) would in turn provide a reasonable estimation of the population served by that particular amenity. In the study by Widener et al. [30], the public transportation service route is used to evaluate the ease at which public transit riders can make a grocery store stop on their way home from work. Using a time constraint of 120 minutes, the existing transportation network, and the transit schedule, the researchers determined a local interaction score for each analysis zone.

Catchment areas can also delineate locations that are ostensibly served by a service facility. This type of information can help determine whether at-risk populations are getting access to health care [31], if patients with mental health issues can get adequate access to specialized doctors, or determine whether there is a differential stress on some services due to higher catchment flows in some others compared to others [32]. Upon inspection, the demand/supply ratios for the catchment areas provide insight into where resources need to be reallocated or where personnel need to be re-assign to better meet the demands of the population in the service area [33]. Analyzing the significance of catchment with respect to access can be determined using a number of geostatistical analysis techniques. These will indicate whether the spatial pattern of access, based on catchment area, is random or if there are clusters of high and low access. For example, after determining the catchment area in their respective study areas, both [30] and [24] used Local Indicators of Spatial Association (LISA) to determine whether there were statistically significant clusters of high and low access, and whether outliers (high access surrounded by low, or vice versa) of access were present across the administrative units of analysis.

This idea of catchment and accessibility can easily be transported to several different services and used to evaluate social capital and resilience. For example, if one aspect of social capital is defined by an individual's ability to receive the social, cultural, or financial support from his/her surrounding community, understanding where the resources are located and how much of that resource is available is critical. Among other things, it provides some sense of demand and supply. As the ratio between demand and supply increases, there is more strain on the system and we would expect less services to be available to those in the catchment area. These could be at risk areas for social unrest due to the lack of services. It may also be areas where individuals begin to look to other illegitimate governments for support.

### 3.4 Information Fusion to Estimate Resilience

Each of the map layers described in Figure 3 corresponds to the density of specific types of structures that indicate localized social capital. The Urban Resilience Ontology describes how each factor contributes to social capital. For example, sectarian services tend to increase homophily for the ingroup and ostracize the outgroup. This leads to high social capital when the local population corresponds to the ingroup, however the same sectarian services would lead to negative social capital when the outgroup occupies the catchment area. The output of the data fusion is a heat map that quantifies the relative social capital for the urban area (Figure 4).

### 3.5 Local Indicators of Spatial Association

Local Indicators of Spatial Association (LISA) are a set of statistical indicators used in exploratory spatial data analysis. The primary purpose of LISA is the identification of spatial autocorrelation across a data set. Although related, LISA statistics are unlike global indicators of autocorrelation, such as the Moran's *I* or Geary's *C* statistics, which only provide an indication of whether autocorrelation is present on a global scale. LISA statistics decompose these global indicators to address the contribution of each individual observation in the data set. That is, significant clusters of observations with similar values can be detected and identified relative to the entire data set [34]. Researchers are able to assess the similarity between different observations based on one variable within a specified distance for a single point. The notion of significant spatial cluster detected at the local level can be conceptualized as a hot/cold spot [35]. The most commonly used LISA statistic is the Local Moran's *I*, formally defined as:

$$I_i = z_i \sum_{i \neq j}^{n} w_{ij} z_j$$

where $z_i$ and $z_j$ are deviations from the mean and $w_{ij}$ is the weights matrix that defines the neighborhood for each observation, $i$. The use of LISA statistics have been broad. Its general applicability makes it highly applicable to a range of research areas where space is seen as a contributing factor to some phenomenon. As an exploratory spatial data analysis technique, it is used to understand the spatial structure of the observations in the data set. For example, LISA statistics have been used to identify where hot spots of future population growth are located and their coincidence with areas of known and predicted environmental stress [36]. Other research has used bivariate LISA statistics to identify significant clusters of social vulnerability and fire risk in the southeastern U.S. [37]. Where the latter is concerned, it is particularly relevant for this study by their use of composite scores describing fire risk and social vulnerability. The fire risk was created by combining spatial data on fire probability, fire suppression effectiveness, and fire behavior and was compared against their constructed spatial indicator of social vulnerability consisting of percent population below poverty, low education attainment, ethnicity, mobile home density, and renter occupied units.

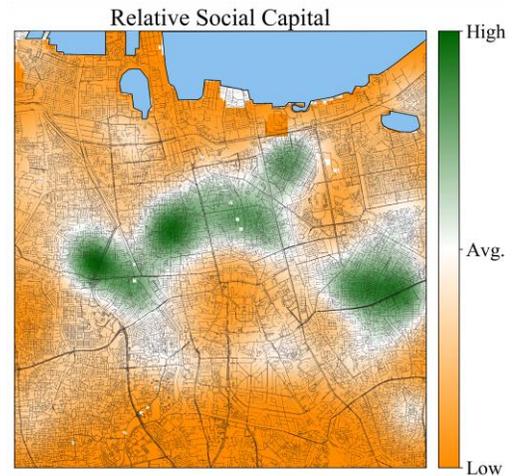

Figure 4: Relative social capital for Jakarta, Indonesia

According to Emrich and Cutter [38], social vulnerability provides "the social context within which such stressors operate and highlight the uneven social capacity for preparedness, response, recovery, and adaptation to environmental threats." As mentioned previously, resilience is the ability to withstand and recover from shocks to the system. Thus, vulnerability will be inversely related to resilience and LISA has been used extensively to identify patterns of vulnerability. In addition to the bivariate LISA analysis for fires detailed above, [39] created a composite score for natural hazard vulnerability (social vulnerability) and used the Local Moran's $I$ to assess whether highly vulnerable counties are clustered in space, and how those clusters vary over time. Interestingly, the lowest socially vulnerable clusters that remained relatively stable across all decades were those with a racially homogeneous population, were affluent, and contained more young people. Other studies using LISA statistics to examine spatial patterns of vulnerability across a landscape include [40] and [41], both of which use nonparametric methods to derive vulnerability scores for natural hazards.

Further still, LISA statistics have been used to identify the movement of populations following a disturbance event. In the case of [42], significant cold spots indicated where in-migration had taken place, hot spots indicated outmigration, and what is referred to as "high-low" areas identified hotspots surrounded by cold spots—the location of the hardest hit areas ("high") and where those who were affected by the disaster left to seek shelter ("low").

Figures 5 and 6 show our Local Moran's $I$ result. Table 1 lists the most statistically significant ($p$-value < 0.001) areas of high and low relative social capital.

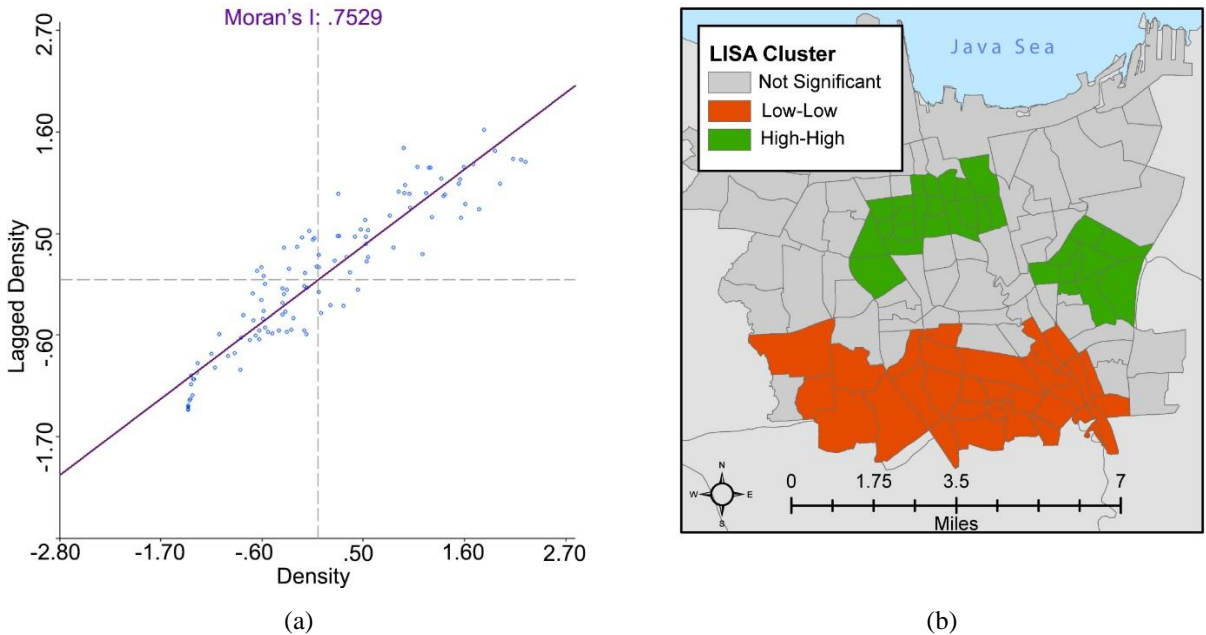

(a)            (b)

Figure 5: (a) Local Moran's $I$ for each neighborhood with statistically significant neighborhoods color-coded. (b) map of statistically significant measurements of Social Capital for Jakarta, Indonesia.

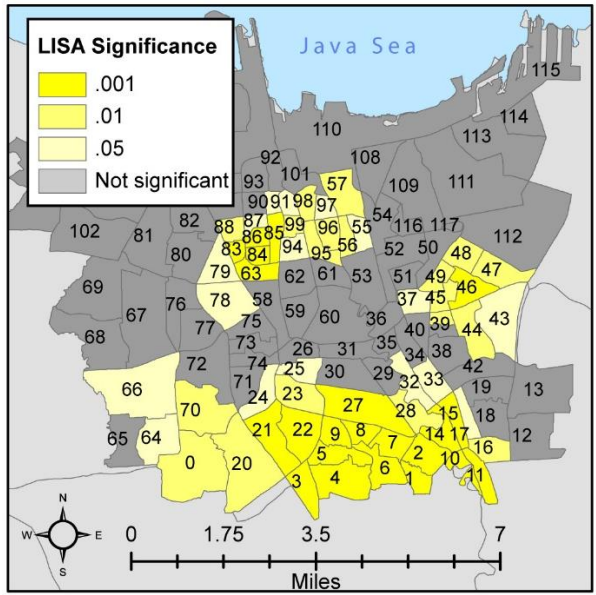

Figure 6: Neighborhoods with statistically significant autocorrelation of high and low relative social capital.

Table 1: Statistically significant high and low relative social capital neighborhoods with $p$-value $\leq 0.001$.

| Stable (High-High) Neighborhood | Feral (Low-Low) Neighborhood |
|---|---|
| 46: Cempaka Baru | 1: Manggarai Selatan |
| 63: Duri Pulo | 2: Manggarai |
| 83: Kalianyar | 3: Karet Semanggi |
| 84: Duri Selatan | 4: Karet Kuningan |
| 85: Tanah Sereal | 5: Karet |
| 86: Duri Utara | 6: Menteng Atas |
|  | 7: Pasar Manggis |
|  | 8: Guntur |

At its core, a LISA statistic provides a method for detecting the presence and strength of spatial association [34]. However, the results of a LISA can spur further investigation into other substantive questions about the data set. As mentioned above, in addition to detecting hot spots and cold spots the LISA provides two other results: low-high and a high-low, where a low value is surrounded by high values or a high value is surrounded by low values, respectively [43]. These unique areas are considered to be outliers and are also tested for significance by the LISA statistics. They provide additional insight into the spatial structure of the data set. For example, in addressing the spatial pattern of homicides in St. Louis, Messner et al. [44], identified significant high-low areas indicating the presence of some sort of barrier that prevents crime from moving from the high area to the surrounding low area, as theories suggest it should. Similar questions were raised with the level of development in the provinces of Turkey [45] and on the spread of disease in India [46]. These patterns can be used to determine the location of hold-outs or particularly well-off area surrounded by impoverished ones as well. Those hold-out areas may indicate high social capital that helps to sustain the population while the specific cultural or social characteristics of the area create a barrier to prevent others from infiltrating the area.

## 4 CONCLUSION

This paper presents a novel approach for fusing geospatial data to estimate urban resilience. We developed a suite of software modules that perform analysis of geospatial indicators of urban resilience from multiple data modalities. Our approach is founded on an ontology that describes geospatial indicators, such as social structures, their catchment areas, and local population density, and how these indicators combine to quantify localized social capital. The ontology distinguishes between bonding and bridging social capital. Increases in localized bonding social capital, coupled with decreases in localized bridging social capital, could be a leading predictor for potential instability or diminishing community resilience. This approach may be useful for identifying hidden linkages that traditionally become apparent only after crisis events. Social capital, measured at fine-grained geospatial scales, may be applicable to disaster relief, humanitarian aid, and the development and planning of "smart cities".

# 5 ACKNOWLEDGEMENTS

The authors would like to thank the SPIE 2019 Signal Processing, Sensor/Information Fusion, and Target Recognition XXVIII conference organizers and the anonymous referees for their valuable comments. Map data copyright by OpenStreetMap contributors and available from https://www.openstreetmap.org. This research was supported by the Defense Advanced Research Projects Agency (DARPA) under Contract No. 140D16318C0063, and this paper has been approved with Distribution Statement "A" (Approved for Public Release, Distribution Unlimited). The views, opinions, and/or findings contained in this document are those of the authors and should not be interpreted as representing the official policies, either expressed or implied, of DARPA or the U.S. Government.